\documentclass[twocolumn,showpacs,aps,prb,floatfix]{revtex4}
\usepackage{graphicx}
\usepackage{subfigure}
\usepackage{epsfig}
\usepackage{psfrag}

\makeatletter

\usepackage{verbatim}

\makeatletter

\input epsf

\def\be{\begin{equation}}
\def\ee{\end{equation}}
\def\ba{\begin{eqnarray}}
\def\ea{\end{eqnarray}}

\makeatother

\begin{document}

\title{Incompatibility of modulated checkerboard patterns with the  neutron scattering resonance peak in cuprate superconductors}                 

\author{D.~X.~Yao  and E.~W.~Carlson}

\affiliation{
Department of Physics, Purdue University, West Lafayette, Indiana  47907,
USA 
}
\date{\today}

\begin{abstract}
  Checkerboard patterns have been proposed to explain the real space
  structure observed in STM experiments on BSCCO and Na-CCOC. However, simple
  checkerboard patterns have low energy incommsensurate (IC) spin peaks rotated
  45 degrees from the direction of the charge IC peaks, contrary to what is
  seen in neutron scattering. Here, we study modulated checkerboard
  patterns which can resolve the low frequency inconsistency. Using spin wave
  theory, we explore the spin response of these superstructures and find that
  the high energy response is inconsistent with neutron scattering results.
In particular, the modulated checkerboard structures are incapable
of supporting the experimentally well-established  resonance peak at ($\pi,\pi$).  
  \end{abstract}
\maketitle

\section{Introduction}
Spin and charge ordering have been topics of great interest in strongly
correlated electronic systems.  A key aspect of many strongly correlated
models is that different terms in the Hamiltonian compete, which introduces a
type of electronic frustration since solutions cannot be found which
simultaneously minimize all terms in the Hamiltonian.  These competing
interactions can lead to spontaneous nanoscale electronic structure.  Indeed,
several locally inhomogenous electronic phases have been proposed, involving
charge order, spin order, and orbital order among others, in strongly
correlated materials such as cuprate superconductors, nickelates, manganites,
and related perovskites.  Charge order is amenable to detection through
probes that directly measure charge degrees of freedom, such as scanning
tunneling microscopy (STM),\cite{hoffman,checknccoc} while the presence of
spin order can be directly detected through neutron
scattering.\cite{tranquadareview,ybco,john,boothroyd03a,boothroyd05,boothroyd06}
Unfortunately, the charge patterns which most naturally explain the STM data
have often been incompatible with the spin patterns which most simply explain
the neutron scattering data.  Part of the challenge has been that materials
which are most amenable to STM studies (i.e. strongly layered materials) are
least amenable to neutron scattering (which requires large crystals), and
vice versa.  Further complicating a clear and consistent interpretation of
the data set as a whole is that while neutron scattering is a bulk probe, STM
is confined to the surface.

\begin{figure}[t]
{\centering
  \subfigure[~Simple Checkerboard\label{fig:simplecheck}]{\resizebox*{0.45\columnwidth}{!}{\includegraphics{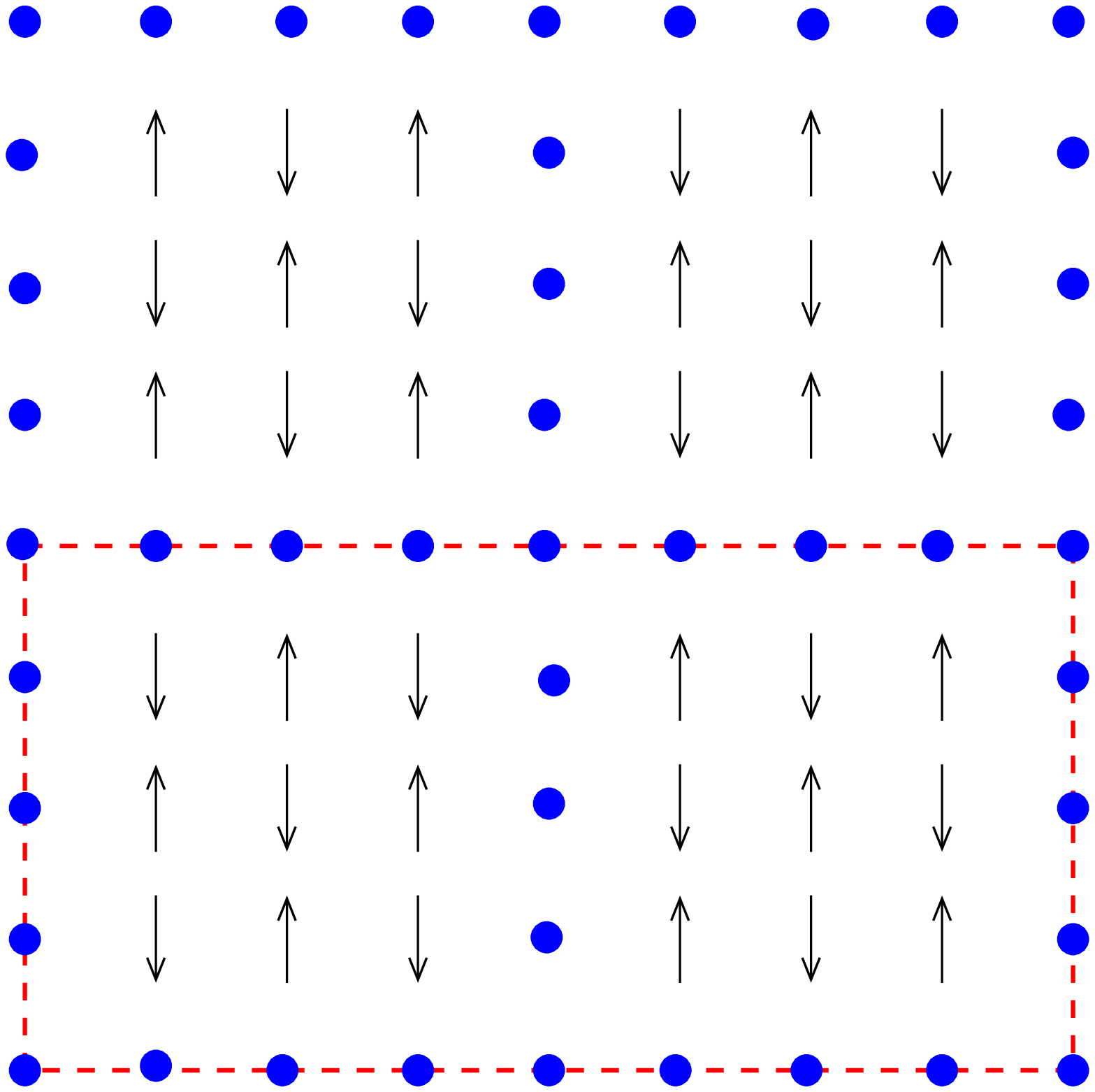}}}
  \hspace{0.05\columnwidth}
  \subfigure[~Noncollinear Checkerboard\label{fig:2q}]{\resizebox*{0.45\columnwidth}{!}{\includegraphics{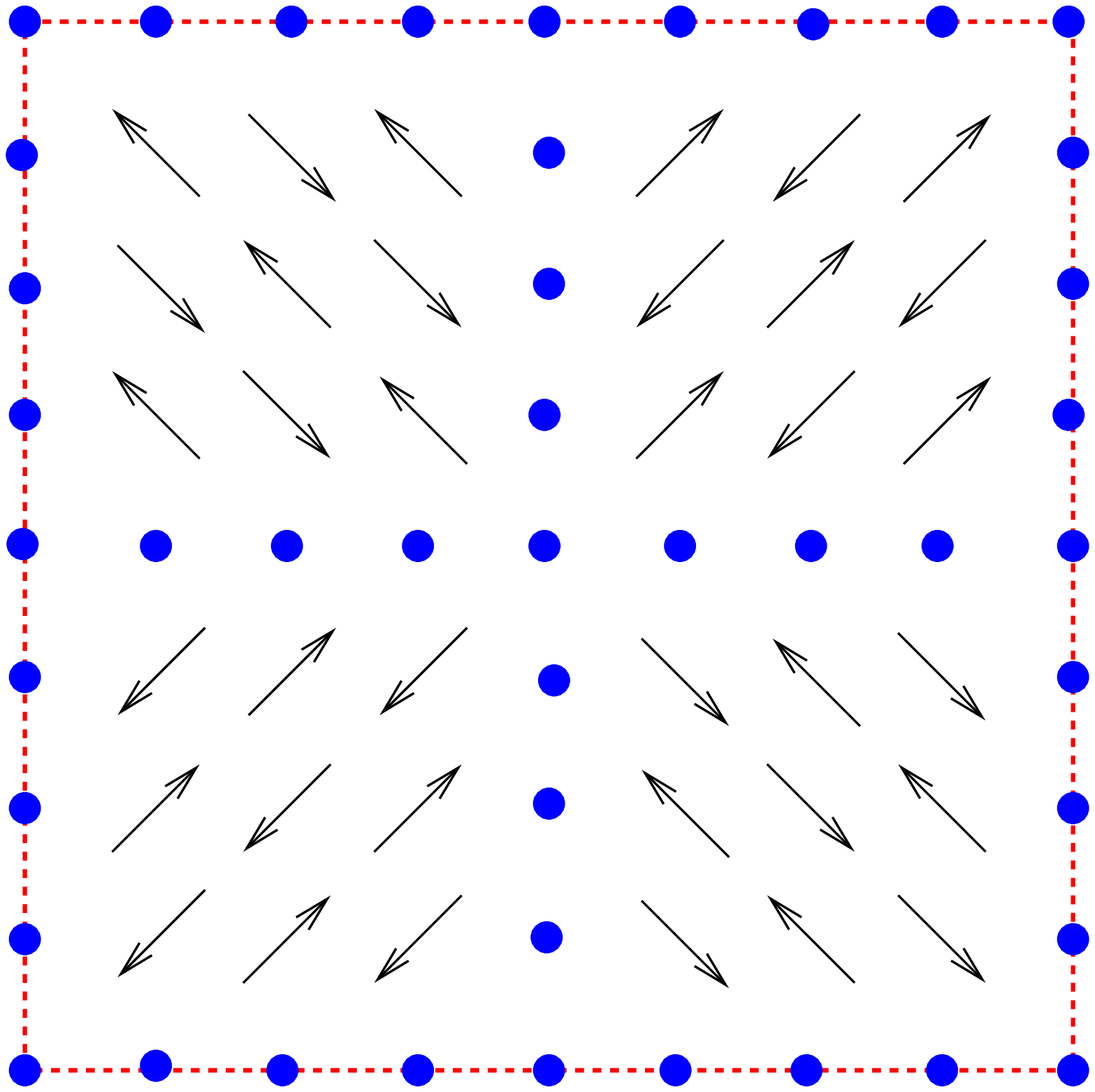}}} \par}
{\centering
  \subfigure[~Modulated Checkerboard\label{fig:acs2}]{\resizebox*{0.45\columnwidth}{!}{\includegraphics{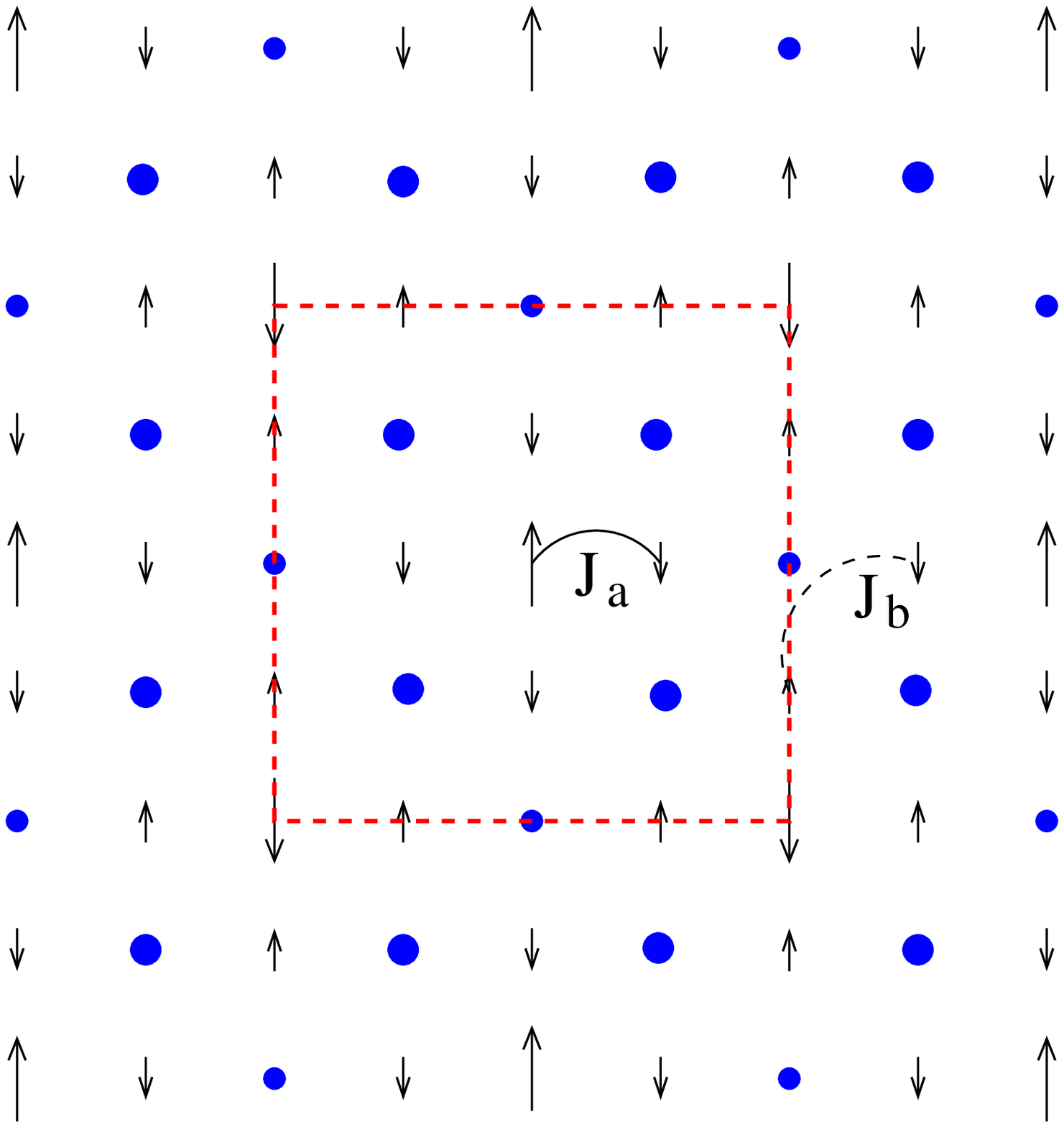}}}
  \hspace{0.05\columnwidth}
  \subfigure[~Modulated Checkerboard\label{fig:acs3}]{\resizebox*{0.45\columnwidth}{!}{\includegraphics{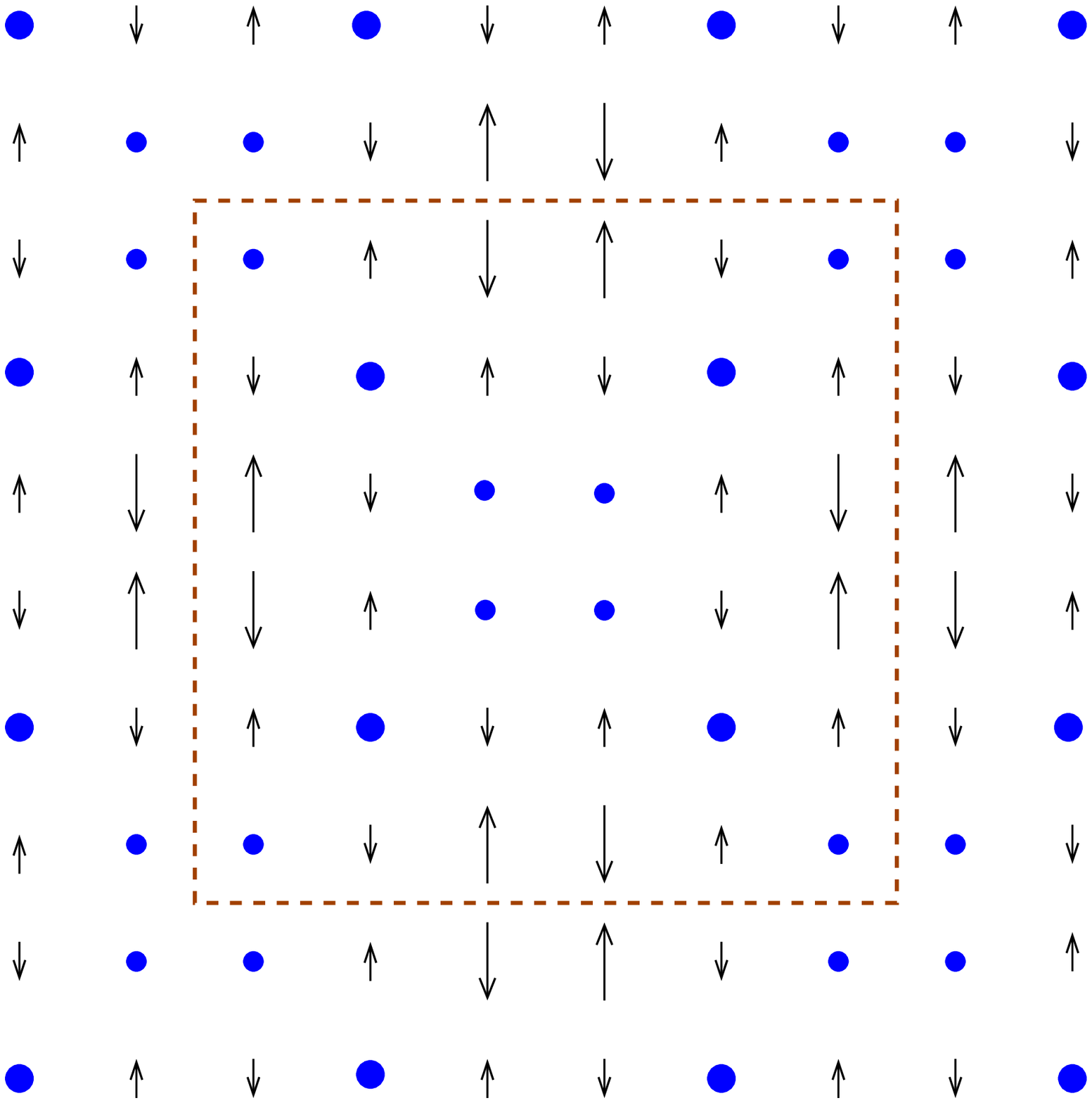}}} \par}
\caption{(Color online) Checkerboard pattern. a) Simple checkerboard with spacing $p=4$. b)
  Noncollinear checkerboard with $p=4$. c) Modulated
  checkerboard pattern with spacing $p=2$. d) Modulated checkerboard pattern
  with spacing $p=3$. The dotted lines represent unit cells.}
\label{lattice}
\end{figure}

Recent experimental advances have made possible the detection of high
energy neutron scattering
spectra~\cite{hayden04,tranquada04a,buyers04}. 
This has led to the discovery that the magnetic excitations 
in several cuprates, {\em i.e.}, La$_{2-x}$Ba$_x$CuO$_4$ (LBCO),
La$_{2-x}$Sr$_x$CuO$_4$ (LSCO), and YBa$_2$Cu$_3$O$_{6+\delta}$ (YBCO),
exhibit universal behavior~\cite{tranquada04a,hayden04,tranquada-schrieffer,tranquada-universal}.
One prominent feature is that at intermediate energies, there is a
resonance peak\cite{keimer95} at $(\pi, \pi)$ formed by the merging of the 
low energy incommensurate response with 
a high energy response whose incommensurate structure is rotated
$45^o$ from that of the low energy excitations. 
The resonance peak is observed in the pseudogap regime ({\em i.e.} the low temperature normal state
of the underdoped cuprates), but sharpens as temperature is lowered into the superconducting state.  
The relation between the resonance peak and
the emergence of superconductivity is still under research.\cite{tranquada06}
The magnetic excitations in these materials have been
explained using stripes, a unidirectional modulation of spin and
charge.~\cite{vojta-ulbricht,uhrig04a,tranquada04a,seibold04,yao06a,yaocarlson06b}.
In these models, the resonance peak is a saddlepoint in the dispersion.


STM can directly detect charge order at the surface. 
Checkerboard patterns (a two-dimensional modulation of charge)
have been proposed to explain the real space
structure observed in STM experiments on
Bi$_2$Sr$_2$CaCu$_2$O$_{8+\delta}$~\cite{hoffman} (BSCCO) and
Ca$_{2-x}$Na$_x$CuO$_2$Cl$_2$~\cite{checknccoc} (Na-CCOC). 
The charge modulations were characterized by checkerboad patterns
with spacing approximately $4a\times 4a$, where $a$ is the lattice spacing. 
While this spacing is in agreement with the  incommensurability
observed in neutron scattering, 
the presence of a true two-dimensional modulation such as a checkerboard
has not been confirmed by neutron scattering or other bulk probes.  

\begin{figure}[!tb]
{\centering 
\resizebox*{0.95\columnwidth}{!}{\includegraphics{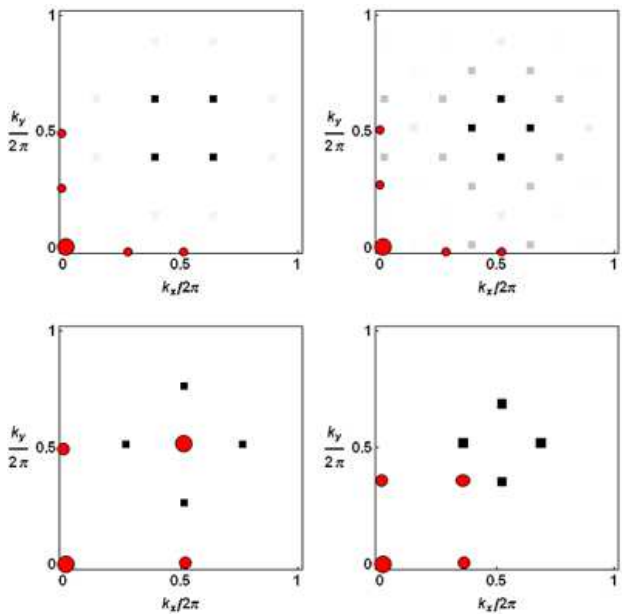}} \par}
\caption{(Color online) Spin (black) and charge (red) expected diffraction peaks of checkerboards in reciprocal lattice space,
  corresponding to the real space patterns of Fig.~\ref{lattice} (a), (b), (c) and (d).}\label{fourier}
\end{figure}

Within microscopic models, either stripes or checkerboards can be
stabilized by tuning parameters, such as the next-nearest neighbor
hopping $t^{\prime}$.\cite{seibold06} The most natural checkerboard
patterns to arise out of microscopic models are ``simple''
checkerboards, such as the one shown in Fig.~\ref{fig:simplecheck}.
Such simple checkerboards are in fact ruled out experimentally,
because the low energy charge peaks are rotated $45^o$ from the low
energy spin peaks, contrary to what is observed in neutron
scattering.\cite{tranquada-schrieffer,vojta-sachdev,andersen03,
yaocarlson06b} Later, modulated checkerboards were
proposed,~\cite{fine04,boothroyd06} 
as shown in Fig.~\ref{fig:acs2} and \ref{fig:acs3}.
Recent experimental work has ruled out the modulated checkerboards in
La$_{1.48}$Nd$_{0.4}$Sr$_{0.12}$CuO$_4$, based on the relative
intensities of the incommensurate (IC) spin peaks in different
magnetic Brillouin zones.\cite{boothroyd06} The authors of
Ref.~\onlinecite{boothroyd06} also proposed a new noncollinear
checkerboard pattern as shown in Fig.~\ref{fig:2q}, which is
consistent with all of the low energy data.  We show here that for
modulated checkerboards there is no possibility of a resonance peak at
($\pi,\pi$), which rules out these structures in all materials where a
resonance peak has been observed.  We further argue that the newly
proposed noncollinear checkerboard also lacks a resonance peak.

\section{Model and Method}
In this paper, we study the magnetic excitations of various checkerboard
patterns. Simple checkerboards of the type shown in
Fig.~\ref{fig:simplecheck} have been studied previously by us and
others\cite{yaocarlson06b,vojta-sachdev,andersen03}.  These types of
patterns, in which the sign of the N\'eel vector changes across each charge
line (whether vertical or horizontal), are always found to have
incommensurate (IC) spin peaks which are rotated $45^o$ from the IC charge
peaks, contrary to what is observed in experiment from STM (which can measure
the charge peaks), and neutron scattering (which can measure the spin peaks,
and sometimes also the charge peaks as well).  Modulated checkerboards, like
those shown in Figs.~\ref{fig:acs2} and \ref{fig:acs3} offer a consistent
description of the low energy data, as does the noncollinear checkerboard
shown in Fig.~\ref{fig:2q}.  Here, we extend our previous work on the
magnetic excitations of stripes and simple checkerboards to the modulated and
noncollinear checkerboard patterns.

We study the spin excitations within the Heisenberg model, 
\begin{equation}
H=   \frac{1}{2}\sum_{i,j} J_{i,j} \mathbf{S}_{i} \cdot \mathbf{S}_{j} 
\label{model}
\end{equation}
where the indices $i$ and $j$ run over all spin sites, and $J_{i,j}$
represents the spin coupling.  We have assumed that the charge degrees
of freedom can be integrated out to produce the effective spin
couplings of the model.  The main effect of the charge degrees of
freedom is to form antiphase domain walls across which the N\'eel
vector of the spins changes sign.  As shown in Fig.~\ref{fig:acs2},
nearest neighbor couplings are $J_a$ (antiferromagnetic), and
next-nearest neigbor couplings across a domain wall are $J_b$ (also
antiferromagnetic).  We make the physically reasonable assumption that
$J_b$ is small compared to $J_a$.  The dotted lines in
Fig.~\ref{lattice} show the unit cells.  Note that in
Figs.~\ref{fig:acs2} and \ref{fig:acs3}, the charge domain walls run
diagonally.  In Fig.~\ref{fig:acs2}, the diagonal spacing between
domain walls is $p=2$ in units of the diagonal spacing $\sqrt{2}a$.
This configuration has $10$ spins in the unit cell.  In
Fig.~\ref{fig:acs3} the diagonal spacing between domain walls is $p=3$
in the same units, and there are $24$ spins in the unit cell.  For
Figs.~\ref{fig:acs2} and \ref{fig:acs3}, the spin ground states shown
are unfrustrated, there is long range magnetic order, and the
elementary excitations can be captured by the spin wave treatment
below.  Note that the noncollinear checkerboard is not a ground state
of this spin model.  We will return to the magnetic excitations of
this state later.

\begin{figure}[!tb]
{\centering 
\resizebox*{0.9\columnwidth}{!}{\includegraphics{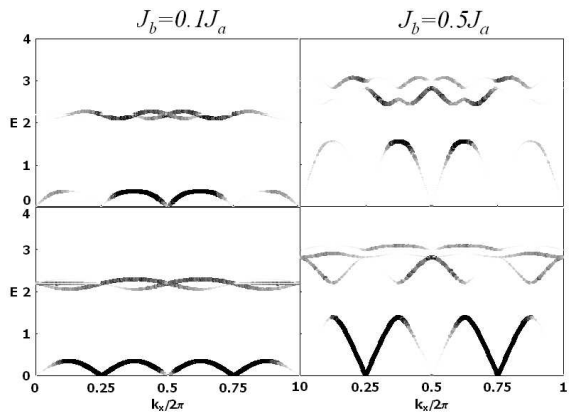}} \par}
\caption{Spin wave dispersion and intensities for a modulated checkerboard pattern
  with spacing $p=2$. The upper panel is along the ($k_x$, $\pi$) direction and
  lower panel is along the diagonal direction ($k_x$, $k_x$). The energy E is in units of $J_aS$.
  }
\label{acs2_disp}
\end{figure}

In order to study the magnetic excitations, we use the spin wave method to
calculate the magnon excitation spectrum and the zero-temperature dynamic structute factor
\begin{equation}
S(\mathbf{k}, \omega)=\sum_f \sum_{i=x,y,z} |\left<f|S^i
(\mathbf{k})|0\right>|^2 \delta (\omega-\omega_f),
\end{equation}
where $\left|0 \right>$ is the magnon vacuum state and $\left|f \right>$
denotes the final state of the spin system with excitation energy $\omega_f$.
$S(\mathbf{k}, \omega)$ is proportional to the expected neutron scattering
intensity. We show two different sizes of spins in Figs.~\ref{fig:acs2} and
\ref{fig:acs3}.  We take the $z$ component of the larger spins to be twice
that of the smaller spins, in order to take into account that the spin
modulation is not a square wave.  However, our major conclusions are
independent of the details of this spin ratio.

\section{Results}
We first discuss the zero-frequency response, shown in Fig.~\ref{fourier}.
The greyscale image shows the expected diffraction peaks
and relative intensities from the spin texture for each panel from Fig.~\ref{lattice}.
The diffraction peaks from the corresponding charge modulation are 
denoted schematically by the red circles, 
for fiduciary points around (0,0).  
For simple checkerboards, 
spin diffraction
peaks are rotated 45 degrees from the direction of charge diffraction
peaks, unlike what is seen in experiment.
For the noncollinear checkerboard of Fig.~\ref{fig:2q}
and the modulated checkerboards of Figs.~\ref{fig:acs2} and \ref{fig:acs3},
the relative orientation of the main spin and charge IC peaks
are consistent with experiment.  
The noncollinear checkerboard gives more satellite peaks
than the modulated checkerboards, even when the spacing
of the noncollinear checkerboard is comparable
to that of the modulated checkerboards, 
which may help distinguish these patterns.
However, the higher harmonic spin peaks get weaker with increasing $p$.
For the experimentally relevant case of  $p=4$,
the main IC peaks are nine times stronger than the next nonzero harmonic peaks,
which may make them difficult to detect. 
Note that for modulated checkerboards of even spacing ($p = $ even),
there is a charge diffraction peak located at ($\pi,\pi$),
contrary to what is observed in experiment. 
The strength of this peak decreases as the charge profile is
made smoother.  

Spin wave dispersions with intensities for the $p=2$ 
modulated checkerboard case are shown in
Fig.~\ref{acs2_disp} at $J_b=0.1J_a$ and $J_b=0.5J_a$ along the $(k_x,\pi)$
and $(k_x, k_x)$ directions.  Note that although zero-frequency weight is
forbidden at $(\pi,\pi)$ because of the presence of antiphase domain walls in
the N\'eel vector, nevertheless, the point $(\pi,\pi)$ is a reciprocal lattice
vector, and so the spin-wave dispersion must approach $\omega \rightarrow 0$
at this wavevector.  There are several defining characteristics of the
resonance peak, but the most well-established is that the mode occurs at
finite frequency.  The reciprocal lattice structure of the modulated
checkerboard patterns therefore forbids the appearance of a resonance peak in
the acoustic band.

For comparison, the simple checkerboard patterns we studied in
Ref.~\onlinecite{yaocarlson06b} are also incapable of supporting a resonance
peak in the acoustic band, but for a different reason.  In these cases,
$(\pi,\pi)$ is {\em not} a reciprocal lattice vector, and so the acoustic
mode at $(\pi,\pi)$ has finite frequency.  However, in the case of simple
checkerboards, the acoustic band reaches a local maximum at $(\pi,\pi)$,
rather than the saddlepoint found in stripe phases.  The saddlepoint
structure has been shown to capture the phenomenology of the resonance
peak,\cite{yaocarlson06b,vojta-sachdev,uhrigsaddlepoint}
including the finite-frequency peak in the integrated spin structure factor
$S(\omega)$, as well as the presence of incommensurate scattering which
smoothly connects to the resonance peak both below and above it in frequency.
By contrast, while the simple checkerboards have a peak in $S(\omega)$ at
finite frequency, there is no weight immediately above the $(\pi,\pi)$ point
in frequency.  This is one of a few reasons why simple checkerboard patterns
have been ruled out.

For the modulated checkerboards studied here, one may consider  
the possibility of the resonance peak appearing
in one of the optical bands.  There is a gap between the acoustic band
and optical bands, similar to the simple checkerboard
configurations.~\cite{yaocarlson06b} The gap is sizable when $J_b$
is small. The acoustic band begins to touch the optical bands at $J_b=J_a$.
For the physically reasonable assumption that $J_b$ is smaller than $J_a$,
any weight appearing in an optical band at $(\pi,\pi)$ is too far removed
from the incommensurate scattering at low frequency to be a 
candidate for the resonance peak.  Furthermore, the lack of a saddlepoint
structure in the optical bands further rules out a resonance peak-like structure
(see Fig.~\ref{acs2_disp}, where 
instead of a saddlepoint, two bands cross at $(\pi,\pi)$).

\begin{figure}[!tb]
{\centering 
\resizebox*{0.8\columnwidth}{!}{\includegraphics{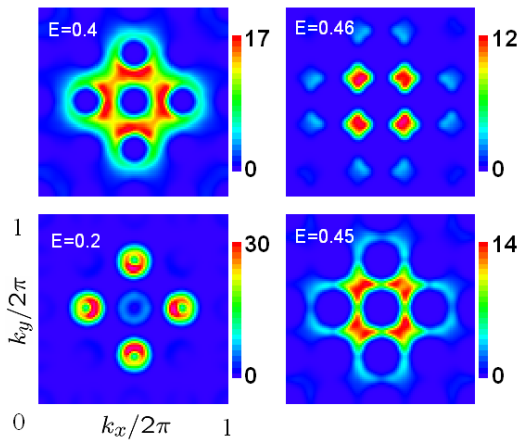}} \par}
\caption{(Color online) Constant energy cuts with windows $0.1J_aS$ for a modulated checkerboard pattern
with diagonal spacing $p=2$ as described in the text.  The coupling
ratio is $J_b/J_a = 0.1$.  The energy $E$ is in units of $J_a S$.
  In each plot, we have integrated over an energy window $\Delta E = \pm 0.1J_a$.}
\label{acs2_cut_0.1}
\end{figure}

Constant energy cut plots are experimentally measurable and useful for
analyzing the microscopic structures.  Fig.~\ref{acs2_cut_0.1} shows
representative constant energy cut plots for the $p=2$ structure of
Fig.~\ref{fig:acs2}, at coupling ratio $J_b/J_a = 0.1$.  For the modulated
checkerboards, it is clear that the direction of the low energy IC spin peaks
are consistent with that observed experimentally for the low energy IC charge
peaks, whereas this was not the case for simple checkerboards.  We find that
although a spin-wave cone must be present at low energy, the intensity of the
spin structure factor is not uniform on the cone, as shown in the $E=0.2$
panel.  As in our previous studies of arrays of antiphase domain walls
(whether in stripe or simple checkerboard patterns),\cite{yaocarlson06b} for
small coupling ratio $J_b/J_a$, the intensity is strongest on the inner
branch of the spin wave cone, {\em i.e.} the side closest to the $(\pi,\pi)$
point.  We take this to be a generic feature of spins which are weakly
coupled across arrays of antiphase domain walls.

Note also the presence of a faint spin wave cone emanating from the
$(\pi,\pi)$ point.  This spin wave cone is required by symmetry, since the
$(\pi,\pi)$ point is a reciprocal lattice vector of the modulated
checkerboard structures. However, due to the antiphase domain walls, weight
is forbidden at zero frequency at $(\pi,\pi)$ and this makes the central cone
quite faint at finite frequency compared to those emanating from the main IC
points.  At intermediate energies ($E=0.4$), the spin wave cones touch each
other.  At higher energies, just below the top of the acoustic band,
incommensurate peaks are once more observed, now rotated to the diagonal
direction, as shown in the $E=0.46$ panel.  Note the complete absence of
weight at the $(\pi,\pi)$ point, precluding a resonance peak from this type
of modulated checkerboard.

\begin{figure}[!tb]
{\centering 
\resizebox*{0.8\columnwidth}{!}{\includegraphics{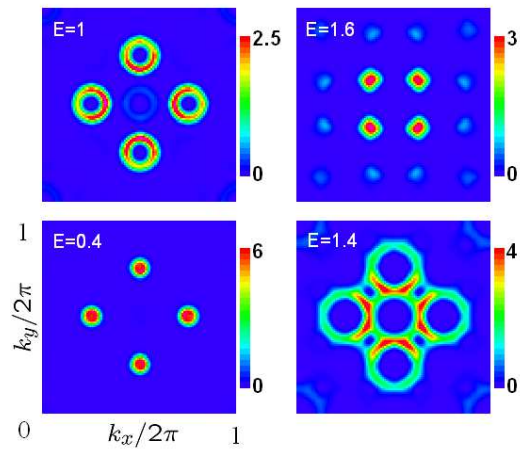}} \par}
\caption{(Color online) Constant energy cuts with windows $0.1J_aS$ for a modulated checkerboard pattern
with diagonal spacing $p=2$ as described in the text.  The coupling
ratio is $J_b/J_a = 0.5$.  The energy $E$ is in units of $J_a S$.
  In each plot, we have integrated over an energy window $\Delta E = \pm 0.1J_a$.
  }\label{acs2_cut_0.5}
\end{figure}
For comparison, in Fig.~\ref{acs2_cut_0.5} we show similar constant energy
cuts, with a stronger coupling ratio $J_b/J_a = 0.5$.  These have a steeper
dispersion (as is evident from Fig.~\ref{acs2_disp}), and the spin wave cones
appear simply as incommensurate peaks in the lowest energy panel, $E=0.4$.
As energy increases, the spin wave cones become visible as shown for $E=1$.  As
with the weaker coupling case, the intensities are not uniform on the spin
wave cone, and the intensities peak on the side facing $(\pi,\pi)$.  Note
that at this energy, the faint spin-wave cone emanating from $(\pi,\pi)$
becomes visible.  At higher energies ($E=1.4$), the spin wave cones merge.
At yet higher energies ($E=1.6$), the top of the acoustic band produces
incommensurate peak structure at finite frequency, with peaks rotated $45^o$
from the low energy IC peaks.  As with the weaker coupling ratio, the
acoustic band is forbidden by symmetry to support any weight at the
$(\pi,\pi)$ point, ruling out these structures as being able to support a
resonance peak.

We now briefly comment on the noncollinear checkerboard proposed in
Ref.~\onlinecite{boothroyd06} and reproduced schematically in our
Fig.~\ref{fig:2q}.  Like the modulated checkerboard patterns, this
pattern has IC spin peaks which are in the same direction as the main
IC charge peaks, consistent with low energy experimental data.  While
modulated checkerboards have been challenged based on the experimental
results of Ref.~\onlinecite{boothroyd06}, the authors were unable to
definitively rule out the noncollinear checkerboard
(``two-$\mathbf{q}$ structure''). We are not able to calculate the
expected magnetic excitation spectrum within the framework of the
current model, since this pattern is not a valid ground state of the
Hamiltonian we consider here (see Eqn.~\ref{model}.)  This does not
preclude it being the ground state of some other model.  What we can
say about this structure is that to the extent that it supports
Goldstone modes, it will have the same limitations of the modulated
checkerboard patterns discussed here.  This is because the $(\pi,\pi)$
point is a reciprocal lattice vector of the noncollinear checkerboard,
and so it is constrained by symmetry to have a spin wave cone
emanating from the $(\pi,\pi)$ point.  This point is also forbidden by
symmetry to have any weight at zero frequency because the pattern has
no net N\'eel vector at $(\pi,\pi)$. Since the intensity of a
Goldstone mode must be continuous in frequency, the spin wave cone
emanating from $(\pi,\pi)$ is constrained to be quite weak.
Therefore, like the modulated checkerboards studied here, the
noncollinear checkerboard will be unable to support a resonance peak
at finite frequency, except perhaps at unphysically high frequencies
in an optical branch.

\section{Conclusions}
In conclusion, we have shown that while modulated checkerboard patterns have
low energy incommensurate charge and spin peaks that are consistent with STM
measurements in BSCCO and Na-CCOC and with neutron scattering measurements in
and neutron scattering experiments on lanthanum cuprates and YBCO, the finite
frequency magnetic excitations of these structures are incompatible with
experimental findings.  In particular, modulated checkerboards are forbidden
by symmetry to have weight at $(\pi,\pi)$ in the acoustic branch, precluding
the possibility of a resonance peak in this branch.  Although optical modes
are not forbidden to have weight at ($\pi,\pi$), the structure of the optical
modes around $(\pi,\pi)$ is incompatible with the phenomenology of the
resonance peak.  
We argue that similar physics constrains the
acoustic branch of the recently proposed noncollinear checkerboard, 
relegating any possibility of a resonance peak to unphysically high energies. 
We conclude that to date
no checkerboard pattern has been proposed which satisfies both the low energy
constraints and the high energy constraints imposed by the current body of
experimental data in cuprate superconductors.

\section*{Acknowledgements}
It is a pleasure to thank D.~K.~Campbell, A.~T.~Boothroyd, and
P.~Abbamonte for helpful discussions.  This work was supported by
Purdue University and Research Corporation (D.X.Y.).  E.W.C. is a
Cottrell Scholar of Research Corporation.

\bibliographystyle{forprb}

\end{document}